\documentclass[reqno]{amsart}

\usepackage{amsmath}
\usepackage{amssymb}
\usepackage{amsthm}
\usepackage{amsfonts}
\usepackage{mathrsfs}
\usepackage[margin=38mm]{geometry}
\usepackage{hyperref}

\usepackage{tabularx}
\usepackage{enumitem} 
\usepackage{braket}
\usepackage{tikz}

\newcommand{\reals}{\mathbb{R}}

\newcommand{\paraa}[1]{\big(#1\big)}

\newcommand{\abs}[1]{\left|#1\right|}
\newcommand{\norm}[1]{\left\lVert #1 \right\rVert}

\newcommand{\diag}{\operatorname{diag}}

\newtheorem{theorem}{Theorem}[section]

\newtheorem{proposition}[theorem]{Proposition}
\theoremstyle{definition}

\theoremstyle{remark}

\numberwithin{equation}{section}

\title{Zero mode surface of Hermitian two-by-two matrices}

\author{Andreas Sykora}
\email{syko@gelbes-sofa.de}

\begin{document}

\begin{abstract}
 We classify the zero mode surfaces, also called Clifford spectra, of Hermitian
 $2\times 2$ matrices. Using the covariance of the Dirac operator under the Euclidean
 group and its invariance under unitary conjugation, together with a singular value
 decomposition, every $m$-tuple of Hermitian $2\times 2$ matrices, after adjoining
 zero matrices if $m<3$, is reduced to the normal form $(b_1\sigma_1,b_2\sigma_2,b_3\sigma_3,0,\dots,0)$ with
 $b_1\geq b_2\geq b_3\geq 0$. Since adding a vanishing matrix only embeds the Clifford
 spectrum into a higher-dimensional space, it suffices to consider three matrices.
 In this normal form the determinant
 of the Dirac operator is given in closed form and its critical points are determined
 explicitly. The zero mode surface consists of one or two points if $b_3=0$. Otherwise
 it consists of two disjoint topological spheres if $b_1>b_2+b_3$, of two topological
 spheres touching at one point if $b_1=b_2+b_3$, and of a single topological sphere if
 $b_1<b_2+b_3$. The topology is thus governed by the triangle inequality for
 $(b_1,b_2,b_3)$, which can be interpreted with Heron's formula. In particular, the
 Clifford spectrum of any tuple of Hermitian $2\times 2$ matrices is never empty.
\end{abstract}

\maketitle

\tableofcontents

\section{Introduction}

For a single Hermitian matrix $X$, the spectrum, i.e. its set of eigenvalues, consists of the real numbers $x$
for which $X-x\mathbf{1}$ is singular. Several inequivalent extensions are available
for a tuple of non-commuting Hermitian matrices, see \cite{CLL26} for an overview.
Here we study the \emph{Clifford spectrum} \cite{Kisil96,Loring15} or \emph{zero mode set}
defined by the zero modes of the Dirac-type operator
\begin{align}\label{intro_dirac}
  D(x)=\sum_{a=1}^m\gamma_a\otimes\paraa{X^a-x^a\mathbf{1}},
\end{align}
where the Hermitian matrices $\gamma_a$ satisfy
$\gamma_a\gamma_b+\gamma_b\gamma_a=2\delta_{ab}\mathbf{1}$. 

There are two physical motivations for studying this set.
In applications to topological materials, a \emph{spectral localizer} combines a
Hamiltonian with position observables in an operator of the form
(\ref{intro_dirac}) \cite{Loring15,LSB17,LSB20}. Its associated invariants remain
constant along invertible deformations. Singular localizers thus identify where an
invariant can change; a singularity does not by itself imply that such a change occurs
\cite{CL22,FG24,CLL26}. This motivates a geometric study of the Clifford spectrum
independently of a particular material model. The applications surveyed in
\cite{CLL26} include topological insulators and superconductors
\cite{HK10,CTSR16,FPL16}, metals and gapless systems \cite{CL22,CCCPLP23}, and
crystalline \cite{CLSB24}, photonic \cite{CL24a} and nonlinear \cite{WLC23} systems.

In string theory, the IKKT and BFSS matrix models \cite{IKKT97,BFSS97} (for reviews see
\cite{Taylor01,Ydri17,Steinacker24}) describe space-time and branes by configurations of
Hermitian matrices, which may be interpreted as quantized embedding functions of a
non-commutative or \emph{fuzzy} space \cite{CDS98}. Well-known examples are the fuzzy
sphere \cite{Hoppe82,Madore92} and fuzzy Riemann surfaces of higher genus
\cite{ABHHS09,Shimada04}. In order to extract the underlying classical geometry from a
given set of matrices, generalized coherent states \cite{Perelomov72} have been used.
They can be defined as quasi-coherent states, i.e.\ ground states of a Laplace-type
operator \cite{Ishiki15,SS16,Steinacker21}, or as zero modes of a Dirac-type operator of
the form (\ref{intro_dirac}) \cite{BD12,BKSY15,KY15}, which can also be motivated by
tachyon condensation \cite{AIMMM18}. For $m=3$ the zero modes of $D(x)$ generically form
a two-dimensional surface in $\reals^3$, the \emph{zero mode surface}, which can be seen
as the shape of the fuzzy membrane described by the matrices. Since the matrices are
finite dimensional, the zero mode surface is the zero set of the polynomial $\det D(x)$
and can be determined without computing the states themselves \cite{BKSY15,Sykora16}.
Based on this observation, a mapping between directed graphs and triples of Hermitian
matrices was introduced in \cite{Sykora16}, which allows to construct fuzzy spaces with
zero mode surfaces of arbitrary genus, see also \cite{Sykora17} for fuzzy spaces with a
classical limit and \cite{SS24} for further examples such as a fuzzy trefoil knot and for
gauge deformations of the fuzzy sphere. The zero mode surface and the Clifford spectrum
are therefore the same object, which is studied in both communities under different names.

The case of $2\times 2$ matrices is the smallest non-trivial case. For $N=1$ or for
commuting matrices the Clifford spectrum simply consists of points, namely of the joint
eigenvalues. The $2\times 2$ case already contains the smallest fuzzy sphere
$X^a=\tfrac12\sigma_a$, whose zero mode surface is a round sphere of radius
$\tfrac12$ \cite{Sykora16,SS24}. Furthermore, in the graph construction of
\cite{Sykora16,SS24} a single edge of a graph corresponds to a $2\times2$ block, the
\emph{basic building block} from which all other fuzzy spaces are composed. Already the
one-parameter family of zero mode surfaces of this building block shows flattened
sphere-like surfaces, round spheres, barbells, pinched surfaces and pairs of disjoint spheres \cite{Sykora16}, and the
merging and splitting of zero mode surfaces observed for larger graphs \cite{SS24}
locally resembles this behaviour. 

On the side of the Clifford spectrum, the $2\times 2$
case is the case in which a non-emptiness conjecture has been proved \cite{CLL26},
and the example $(\tfrac12\sigma_1,\sigma_2,\tfrac12\sigma_3)$ shows that the Clifford
spectrum does not need to be a manifold, since it is a rotated lemniscate of Bernoulli
\cite{DLS22}. After removing the trace parts and diagonalizing the first matrix, \cite{CLL26}
works with seven real coefficients. Its Proposition~2.3 expresses the determinant
through a real symmetric quadratic-form matrix, and its non-emptiness proof uses the
largest eigenvalue of that matrix. Its geometric discussion relates the spectrum to
intersections of Cassini surfaces with quadrics and describes connected, touching and
disconnected configurations, including point degenerations. Coordinate rotations
also occur in that discussion.

The additional reduction presented herein uses orthogonal mixing of the entire tuple together with
unitary conjugation. A singular value decomposition then gives the normal form
$(b_1\sigma_1,b_2\sigma_2,b_3\sigma_3,0,\dots,0)$, with
$b_1\geq b_2\geq b_3\geq0$, after adjoining zero matrices if $m<3$.
For nonzero traceless tuples, removing an overall scale leaves two parameters.
We relate this normal form explicitly to the determinant in \cite{CLL26}, give
singular-value criteria for each topology, and prove the classification by a radial
parametrization of the zero set. For positive singular values the transition is
governed by the triangle inequality, with a geometric interpretation through Heron's
formula. Finally, a zero-matrix extension argument carries the result from triples
to tuples of arbitrary length. Thus the contribution is an invariant parametrization
and a direct topological proof refining the earlier geometric description, together
with this extension in the tuple length.

The paper is organized as follows. In Section \ref{sec_sym} we discuss the symmetries
of the Dirac operator and of its determinant. In Section \ref{sec_class} we derive the
normal form for $m$ Hermitian $2\times 2$ matrices, classify the corresponding zero
mode surfaces and describe the moduli space.
Section \ref{sec_concl} contains conclusions and an outlook.

\section{Symmetries}\label{sec_sym}

Let $X^a$ be $m$ Hermitian $N\times N$ matrices. The Dirac operator
\begin{align}
  D(x) = \gamma_a \otimes \paraa{X^a - x^a}  
\end{align}
is covariant with respect to rotations and shifts $X^a\mapsto R^a{}_bX^b+c^a\mathbf{1}, x^a\mapsto R^a{}_bx^b+c^a$
\begin{align}
  D(x) \mapsto \gamma_a \otimes R^a{}_b \paraa{ X^b - x^b}  = S(R)^{-1}\gamma_b S(R) \otimes \paraa{ X^b - x^b} 
    = \paraa{ S(R)^{-1}\otimes 1} D(x) \paraa{ S(R)\otimes 1}
\end{align}
where $R^a{}_b$ is a rotation matrix in $SO(m)$, $c^a$ is a vector in $\reals^m$
and $S(R)$ is an element of the spin group $\mathrm{Spin}(m)$ with $S(R)\gamma^aS(R)^{-1}=R^a{}_b\gamma^b$.
(The shifts cancel in $X^a-x^a$, so that $D(x)$ is invariant under them.)

Additionally, the Dirac operator is covariant with respect to unitary transformations
$X^a \mapsto U X^a U^\dagger$
\begin{align}
  D(x) \mapsto \gamma_a \otimes \paraa{U X^a U^\dagger - x^a}  
    = \gamma_a \otimes U \paraa{X^a - x^a}  U^\dagger 
    = \paraa{1\otimes U} D(x) \paraa{1\otimes U^\dagger} 
\end{align}
where $U$ is a unitary matrix in $U(N)$. (Note that conjugation with a unitary matrix
preserves the property of being Hermitian. )

For a square matrix $M$ and an invertible matrix $G$ of the same size,
$\det(GMG^{-1})=\det M$. Thus the determinant of the Dirac operator $\det D(x)$
is invariant under the transformations above. 

Up to now we have only treated rotations without reflections. For even $m$
we consider conjugation with $\gamma_1$, which implements $R=\diag(1,-1,\dots,-1)$, an orientation-reversing
transformation. The determinant of the Dirac operator $\det D(x)$ is invariant for
even $m$, since the Dirac operator has even matrix size.
When $m$ is odd, every element $R\in O(m)$ is a rotation in combination with the 
special reflection $R=-\mathbf{1}$, which sends $D(x)$ to $-D(x)$. In this case
the determinant can take up a sign.

\medskip
For $m=3$ the set of $x\in \reals^3$ with $\det D(x)=0$ is called \emph{zero mode surface}
\cite{Sykora16,SS24}
and more generally for $x\in\reals^m$ \emph{zero mode set} or \emph{Clifford spectrum} \cite{CLL26}.
We denote it by $\Lambda(X^1,\dots,X^m)$. 
Since $\det -D(x)=0$ also when $\det D(x)=0$, the zero mode set
is also invariant with respect to reflections.

In summary, we arrive at 
\begin{proposition}
  Let $X^1,\dots,X^m$ be Hermitian $N\times N$ matrices. The
  zero mode set or Clifford spectrum $\Lambda(X^1,\dots,X^m)$
  is invariant with respect to unitary conjugations and
  transforms covariantly with respect to the Euclidean group $E(m)$, i.e.\ when the
  matrices are rotated, reflected and shifted, the Clifford spectrum is rotated, reflected and shifted
  correspondingly. 
\end{proposition}

\medskip
The following proposition answers the question what happens to the Clifford spectrum,
when a vanishing matrix is added to a tuple. This is important for the classification below,
where for $m>3$ matrices, $m-3$ matrices are rotated to $0$.

\begin{proposition}\label{prop_zero_matrix}
  Let $X^1,\dots,X^m$ be Hermitian $N\times N$ matrices and $X^{m+1}=0$. Then
  \begin{align}
    \Lambda(X^1,\dots,X^m,0) = \Lambda(X^1,\dots,X^m)\times\{0\} \subset \reals^{m+1}.
  \end{align}
\end{proposition}

\begin{proof}
  Let $\Gamma_1,\dots,\Gamma_{m+1}$ be Hermitian generators of an irreducible representation
  of the Clifford algebra of $\reals^{m+1}$, i.e.\ $\Gamma_a\Gamma_b+\Gamma_b\Gamma_a=2\delta_{ab}$.
  For $(x,t)\in\reals^m\times\reals$ the Dirac operator of the extended tuple is
  \begin{align}
    L(x,t) = \tilde D(x) - t\,\Gamma_{m+1}\otimes\mathbf{1}, \qquad
    \tilde D(x) = \sum_{a=1}^m \Gamma_a\otimes\paraa{X^a-x^a}.
  \end{align}
  Since $\Gamma_{m+1}$ anticommutes with $\Gamma_1,\dots,\Gamma_m$, the operator $\tilde D(x)$
  anticommutes with $\Gamma_{m+1}\otimes\mathbf{1}$, and therefore
  \begin{align}
    L(x,t)^2 = \tilde D(x)^2 + t^2\,\mathbf{1}.
  \end{align}
  Since $\tilde D(x)$ is Hermitian, $L(x,t)$ is singular if and only if $t=0$ and $\tilde D(x)$
  is singular.

  It remains to show that $\tilde D(x)$ is singular if and only if $D(x)$ is singular.
  The matrices $\Gamma_1,\dots,\Gamma_m$ generate a representation of the Clifford algebra of
  $\reals^m$, which decomposes into a direct sum of irreducible representations, and
  $\det\tilde D(x)$ is the product of the corresponding determinants. If $m$ is even, every
  irreducible representation is equivalent to the one given by the $\gamma_a$, so that the
  corresponding determinant equals $\det D(x)$. If $m$ is odd, there are two inequivalent
  irreducible representations, which can be chosen as $\gamma_a$ and $-\gamma_a$; the latter
  leads to $-D(x)$, which has the same zero set as $D(x)$. Hence $\det\tilde D(x)=0$ if and
  only if $\det D(x)=0$.
\end{proof}

Applying the proposition repeatedly, we obtain
$\Lambda(X^1,\dots,X^m,0,\dots,0)=\Lambda(X^1,\dots,X^m)\times\{0\}^k$ for $k$ vanishing matrices.

\section{Classification of the Clifford spectrum of $m$ Hermitian $2\times 2$-matrices}\label{sec_class}

We assume that the $X^a$ are $m$ Hermitian $2\times 2$ matrices, i.e. $N=2$.
We can decompose every $2\times 2$-matrix into a linear
combination of the identity matrix $\sigma_0 =\mathbf{1}$ and the Pauli matrices $\sigma_i$.
\begin{align} \label{pauli_decomp}
   X^a = x^a_0 \sigma_0 +x^a_i\sigma_i
\end{align}
$U(2)$ acts on the Pauli matrices as a rotation $U\sigma_i U^\dagger=R(U)_i{}^j\sigma_j$ with $R(U)\in SO(3)$
and therefore
\begin{align}
  U  X^a U^\dagger= x^a_0 \sigma_0 + x^a_iR(U)_i{}^j\sigma_j
\end{align}
$E(m)$ acts by a shift $s^a$ and an orthogonal transformation $R(E)$ on the other index
\begin{align}
  X^a \mapsto  R(E)^a{}_b(x^b_0 +s^b) \sigma_0 + R(E)^a{}_b x^b_i R(U)_i{}^j\sigma_j
\end{align}
By a shift $s^a=-x^a_0$, we can eliminate the first term. The real $m\times 3$ matrix $M=(x^b_i)$
in the second term transforms as $M\mapsto R(E)\,M\,R(U)$ and can be diagonalized with a singular value decomposition.

In general, with a singular value decomposition
every real $m\times n$-matrix $M$ can be decomposed into $M=R_1DR_2^T$,
where $R_1$ is an orthogonal $m\times m$-matrix, $R_2$ is an orthogonal $n\times n$-matrix, 
and $D$ is an $m \times n$ diagonal matrix with non-negative entries.

In the present case $n=3$, i.e. three Pauli matrices, the right factor can be chosen in $SO(3)$ without
changing the non-negative diagonal entries of $D$. To see this, suppose first that
$m\geq3$. Set $\varepsilon=\det R_2$, $K=\diag(1,1,\varepsilon)$, and let $J$ be the
$m\times m$ diagonal matrix whose third entry is $\varepsilon$ and whose other
entries are $1$. Since $JDK=D$, replacing $R_1$ by $R_1J$ and $R_2$ by $R_2K$
leaves $M=R_1DR_2^T$ unchanged and makes the right factor a proper rotation.
For $m<3$, the third column of $D$ vanishes, so the same adjustment of $R_2$ can
be made without changing $R_1$.
We henceforth use these adjusted factors. The double cover $SU(2)\to SO(3)$
provides a unitary $U$ with $R(U)=R_2$. Taking $R(E)=R_1^T$ then gives
$R(E)MR(U)=D$. The left factor may be any orthogonal matrix because the Clifford
spectrum is covariant under the full group $O(m)$.

In the present case, for $m\geq 3$ we can transform the arbitrary Hermitian matrices $X^a$
into $b_a\sigma_a$ (no summation) for $a=1,2,3$ and $X^a=0$ for $a>3$,
where the $b_a$ are the singular values of the matrix $M=(x^a_i)$.
For $m<3$ the matrix $M$ has only $m$ singular values, i.e.\ $X^a=b_a\sigma_a$ for $a\leq m$.

We can formulate the following proposition.
\begin{proposition} \label{prop_b1b2b3rep}
  With respect to the $E(m) \times U(2)$-symmetry,
  a set of $m\geq 3$ Hermitian $2\times 2$-matrices $\{ X^a \}$
  is equivalent to $\{ b_1 \sigma_1, b_2 \sigma_2, b_3 \sigma_3, 0, \dots, 0 \}$,
  where $b_a$ are three non-negative parameters and $\sigma_a$ are the Pauli matrices.
  For $m<3$ it is equivalent to $\{ b_1 \sigma_1,\dots, b_m\sigma_m \}$.
\end{proposition}

\medskip
Permuting singular directions simultaneously on the left and the right of $D$
permutes its diagonal entries. If the resulting right orthogonal factor has negative
determinant, the compensating sign adjustment above restores a factor in $SO(3)$.
Thus the tuples $\{b_{P(i)}\sigma_i\}$ and $\{b_i\sigma_i\}$ (no summation) are
equivalent under the stated symmetry action for every permutation $P$.
Geometrically, the corresponding coordinate axes are exchanged.

Thus, it is possible to assume
\begin{align} \label{mod_sp_cone}
  b_1 \geq b_2 \geq b_3 \geq 0
\end{align}

The moduli space of possible $b_i$ reduces to a cone in $\reals^3$
between rays starting at the origin and going through the points
$(1,0,0)$, $(1,1,0)$ and $(1,1,1)$.

\subsection{Zero mode surface}

The non-emptiness theorem and geometric analysis in \cite{CLL26} provide the
starting point for this classification. Here we use the singular value decomposition to express
its determinant in invariant parameters and to give a direct proof of the topology
of every case. Unlike the coordinate rotations used in its geometric discussion,
our reduction combines orthogonal mixing of the entire triple with unitary
conjugation of the matrices.

We determine the zero mode surface of the Dirac operator $D(x) =\sigma_a\otimes \paraa{X^a-x^a}$
for three $2 \times 2$-matrices. The canonical form of the three matrices is
\begin{align}
  X=X^1 = b \sigma_1, \qquad Y=X^2 = c \sigma_2, \qquad Z=X^3 = d \sigma_3
\end{align}
where we have set $b=b_1, c= b_2, d = b_3$.

Explicitly, the localized Dirac operator $D(x) = \sigma_a (X^a-x^a)$ becomes
\begin{align}
  D = \begin{pmatrix}
   d - z     & 0                   & -x + iy         & b - c \\
    0                   & -d - z     & b + c & -x + iy         \\
    -x - iy         & b + c &  - d+ z      & 0                   \\
    b - c & -x - iy         & 0                   & d + z
\end{pmatrix}
\end{align}
resulting in the determinant
\begin{align}
 F(x,y,z) = \det D \label{eq_det_normal} 
  & = \paraa{ b^2 + c^2 + d^2 + x^2 + y^2 + z^2 }^2 
  - 4\paraa{ b^2 x^2 + c^2 y^2 + d^2 z^2} \\
  & - 4 \paraa{b^2c^2 +b^2d^2+c^2d^2 }
\end{align}

Note that the expression for $\det D$ is completely symmetric with respect to
permuting $x,y,z$ and $b,c,d$ in the same way. Reason for this is that the
shape of the zero mode surface does not change, but is solely rotated.

Let us first consider the case, when $d=0$ and $b \geq c \geq 0$.
The determinant reduces to
\begin{align}
 \det D & = \paraa{ b^2 + c^2  + x^2 + y^2 + z^2 }^2 
  - 4\paraa{ b^2 x^2 + c^2 y^2 + b^2c^2   } \\
  & = \paraa{x^2 + y^2 + z^2 + c^2 - b^2 }^2 
  + 4\paraa{ b^2 - c^2 } y^2 + 4 b^2z^2   
\end{align}
Since $b\geq c$ also the second term is non-negative, and $\det D=0$ only holds, when
\begin{align}
    x^2 +y^2 + z^2 +c^2 - b^2=0, \qquad \paraa{ b^2 - c^2 } y =0, \qquad b z = 0
\end{align}

If $b=c$, it follows that $x^2+y^2+z^2=0$ and $(x,y,z)=(0,0,0)$ is the
only single solution.
If $b\neq c$, then $z=0$ and $y=0$ and therefore $x^2=b^2-c^2 > 0$.
In summary, when $d=0$ and $b\geq c$, the zero mode surface consists of
the one or two points
\begin{align} \label{xpm_0}
     x_{\pm,d=0} = \paraa{\pm\sqrt{b^2-c^2}, 0, 0 }
\end{align}

\medskip
From now, we assume that $b\geq c \geq d > 0$, i.e. that all the parameters
are greater than zero.

Critical points of $F=\det D$ are, where the partial derivatives 
\begin{align}
 \partial_x F & =  4x \paraa{ - b^2 + c^2 + d^2 + x^2 + y^2 + z^2 } \\
 \partial_y F & =  4y \paraa{  b^2 - c^2 + d^2 + x^2 + y^2 + z^2 } \\
 \partial_z F & =  4z \paraa{  b^2 + c^2 - d^2 + x^2 + y^2 + z^2 } 
\end{align}
vanish. Since $b\geq c\geq d>0$, the brackets in $\partial_yF$ and $\partial_zF$ are positive,
so that $y=z=0$ at every critical point. Thus, critical points are only the origin $(0,0,0)$ and the points
$(x_\pm, 0, 0)$ with $x_\pm = \pm\sqrt{b^2 -c^2-d^2}$, when $b^2>c^2+d^2$.
In particular, this means that the critical points are solely on the $x$-axis.

The Hessian at the origin is 
\begin{align}
  H\big|_0 = 4\begin{pmatrix} -b^2+c^2+d^2 & 0 & 0 \\ 0 & b^2-c^2+d^2 & 0 \\ 0 & 0 & b^2+c^2-d^2 \end{pmatrix}
\end{align}
The value of $\det D$ at the origin is
\begin{align} \label{origin_val}
   D_0 = \det D(0,0,0) & = b^4 + c^4 + d^4 -2 b^2c^2 - 2b^2d^2 -2c^2d^2  \\
   &= (b-c-d)(b-c+d)(b+c-d)(b+c+d)
\end{align}
When $b,c,d$ are considered as the side
lengths of a triangle, the area $A$ of the triangle satisfies 
\begin{align} \label{heron}
  16A^2=(b+c+d)(-b+c+d)(b-c+d)(b+c-d)
\end{align}  
according to Heron's formula.
It follows that $ D_0 =-16A^2$ and $D_0$
is negative, when a triangle with side lengths $b,c,d$ can be formed,
vanishes at the degenerate triangle $b=c+d$
and is positive when $b,c,d$ fail the triangle inequality.

The Hessian at the two other points on the $x$-axis with $x_\pm$ is
\begin{align}
  H\big|_{x\text{-axis}} = 8\begin{pmatrix} b^2-c^2-d^2 & 0 & 0 \\ 0 & b^2-c^2 & 0 \\ 0 & 0 & b^2-d^2 \end{pmatrix}
\end{align}
The value of $\det D$ at these points is
\begin{align} 
   D_\pm = \det D(x_\pm,0,0) =   - 4 c^2d^2
\end{align}
which is negative, if $c$ and $d$ are both positive, or zero,
when one of them vanishes.

Since $F(x,y,z)\to+\infty$ as $\abs{(x,y,z)}\to\infty$ and $F$ is continuous,
the zero mode surface $F=0$ is non-empty as soon as $F$ takes a negative value somewhere.
We already know that when $d=0$, the zero mode surfaces consists of at least
one point (\ref{xpm_0}). Otherwise $c\geq d$, $c$ and $d$ are both nonzero
and $D_\pm$ is negative at $x_\pm$, thus the zero mode surface $F=0$
is always non-empty in accordance with \cite{CLL26}.

\medskip
By the implicit function theorem for a function $f$, if $f(p)=0$ and
$\nabla f(p)\neq 0$ at a point $p$, then the zero set $f=0$
is a smooth $2$-dimensional surface near $p$. In particular $p$
cannot be an isolated point of the zero set. 

The zero mode surface $F=0$ is therefore a smooth $2$-dimensional surface,
except when critical points of $F$ lie directly on it, i.e. when
$F(p)=0$ and $\nabla F(p)=0$. For $d>0$, this is only the case at the origin,
when $D_0=0$, i.e. in the degenerate triangle case $b=c+d$
(note that $F(x_\pm,0,0)=D_\pm<0$ then).

The overall shape of the zero mode surface follows from
the next proposition, which determines the shapes of the curves,
which are generated by intersecting the zero mode surface with the
planes orthogonal to the $x$-axis.

\begin{proposition}\label{prop_slice}
  Assume $b\ge c\ge d>0$ and let $x_0\in\reals$. Set
  \begin{align}
    R_\pm=b^2-(c\pm d)^2, \qquad
    C(x_0)=F(x_0,0,0)=\paraa{x_0^2-R_-}\paraa{x_0^2-R_+}.
  \end{align}
  Then the slice $\{(y,z)\in\reals^2:\ F(x_0,y,z)=0\}$ of the zero mode surface
  with the plane $x=x_0$ is
  \begin{enumerate}[label=\textup{(\roman*)}]
    \item empty, if $C(x_0)>0$;
    \item the single point $(0,0)$, if $C(x_0)=0$;
    \item a smooth simple closed curve, if $C(x_0)<0$. It is symmetric under
      $y\mapsto-y$ and $z\mapsto-z$ and star-shaped with respect to $(0,0)$.
      Its extent along the $y$- and $z$-axes is given by the semi-axes
      \begin{align}
        y_{\max}^2=-\alpha+\sqrt{\alpha^2-C}, \qquad
        z_{\max}^2=-\beta+\sqrt{\beta^2-C},
      \end{align}
      where $\alpha=b^2-c^2+d^2+x_0^2$ and $\beta=b^2+c^2-d^2+x_0^2$.
      One has $y_{\max}\ge z_{\max}$, with equality if and only if $c=d$, in
      which case the curve is a circle.
  \end{enumerate}
  Moreover, $C(x_0)<0$ holds exactly for $R_+<x_0^2<R_-$, so that the
  non-empty slices are the following.
  \begin{enumerate}[label=\textup{(\alph*)}]
    \item If $b \leq c+d$: -$\sqrt{R_-} \leq x_0 \leq \sqrt{R_-}$.
    \item If $b>c+d$: $-\sqrt{R_-} \leq x_0 \leq -\sqrt{R_+}$ and $\sqrt{R_+} \leq x_0 \leq \sqrt{R_-}$
  \end{enumerate}
  At the end of the intervals, the curves degenerate to the point $(y,z)=(0,0)$ and additionally,
  if $b=c+d$, at $x_0=0$.
\end{proposition}

\begin{proof}
  Put $S=b^2+c^2+d^2$ and $r^2=y^2+z^2$. By (\ref{eq_det_normal}),
  \begin{align}
    F(x_0,y,z)=\paraa{r^2+S+x_0^2}^2-4c^2y^2-4d^2z^2-4b^2x_0^2-4\paraa{b^2c^2+b^2d^2+c^2d^2}.
  \end{align}
  Expanding and collecting terms gives $F(x_0,y,z)=G(y,z)$ with
  \begin{align}
    G(y,z)=r^4+2\alpha y^2+2\beta z^2+C,
  \end{align}
  where $\alpha=S+x_0^2-2c^2$, $\beta=S+x_0^2-2d^2$ and
  $C=\paraa{S+x_0^2}^2-4b^2x_0^2-4\paraa{b^2c^2+b^2d^2+c^2d^2}$. Evaluating at
  $y=z=0$ shows $C=F(x_0,0,0)$. Using $d_0=S^2-4\paraa{b^2c^2+b^2d^2+c^2d^2}$ from
  (\ref{origin_val}), we get
  \begin{align}
    C=x_0^4+2\paraa{c^2+d^2-b^2}x_0^2+d_0=\paraa{x_0^2-R_-}\paraa{x_0^2-R_+},
  \end{align}
  since the roots of this quadratic in $x_0^2$ are $b^2-(c\mp d)^2$.
  Because $b\ge c\ge d>0$, we have
  \begin{align}
    \alpha=b^2-c^2+d^2+x_0^2>0, \qquad
    \beta=\alpha+2\paraa{c^2-d^2}\ge\alpha>0.
  \end{align}

  (i), (ii) If $C\ge0$, every term of $G$ except possibly $C$ is non-negative.
  Hence $G=0$ forces $r=0$ and $C=0$. For $C>0$ there is no solution. For $C=0$
  the only solution is $(y,z)=(0,0)$.

  (iii) Let $C<0$ and write $y=r\cos\theta$, $z=r\sin\theta$ with $r\ge0$. Then
  $G=0$ is equivalent to
  \begin{align}
    r^4+2g(\theta)r^2+C=0, \qquad g(\theta)=\alpha\cos^2\theta+\beta\sin^2\theta>0.
  \end{align}
  This quadratic in $r^2$ has product of roots $C<0$, hence exactly one positive
  root
  \begin{align}
    r^2(\theta)=-g(\theta)+\sqrt{g(\theta)^2-C},
  \end{align}
  which depends continuously on $\theta$. The slice is therefore the image of
  the circle under the injective continuous map $\theta\mapsto r(\theta)(\cos\theta,\sin\theta)$,
  so it is a simple closed curve, star-shaped with respect to $(0,0)$. Since $g$
  is invariant under $\theta\mapsto-\theta$ and $\theta\mapsto\pi-\theta$,
  the curve is symmetric under $z\mapsto-z$ and $y\mapsto-y$. Smoothness follows from
  \begin{align}
    \partial_yG=4y\paraa{r^2+\alpha}, \qquad \partial_zG=4z\paraa{r^2+\beta},
  \end{align}
  which vanish simultaneously only at $y=z=0$, where $G=C\neq0$. Hence
  $\nabla G\ne0$ on the curve, and the curve is smooth by the implicit function theorem.

  The semi-axes are $r^2(0)$ and $r^2(\pi/2)$, i.e.\ the formulas stated in (iii).
  The function $\varphi(g)=-g+\sqrt{g^2-C}$ is strictly decreasing on $g>0$, since
  $\varphi'(g)=-1+g/\sqrt{g^2-C}<0$ for $C<0$. As $\alpha\le\beta$, this gives
  $y_{\max}\ge z_{\max}$, with equality if and only if $\alpha=\beta$, i.e.\ $c=d$.
  In that case $g$ is constant, $r$ does not depend on $\theta$, and the curve
  is a circle.

  Finally, $C=\paraa{x_0^2-R_-}\paraa{x_0^2-R_+}<0$ if and only if $R_+<x_0^2<R_-$ because
  $R_+<R_-$. Since $0\le c-d<c\le b$, we have $R_->0$. If $b \leq c+d$, then $R_+\leq 0\le x_0^2$
  and the condition reads $\abs{x_0}<\sqrt{R_-}$, which is (a). If $b>c+d$,
  then $R_+>0$ and the condition reads $\sqrt{R_+}<\abs{x_0}<\sqrt{R_-}$, which is (b).

  For $\abs{x_0}>\sqrt{R_-}$ we have $C>0$, so the slice is empty. At
  $x_0^2\in\{R_-,R_+\}$ with $x_0^2\ge0$ we have $C=0$, so by (ii) the slice is the
  point $(0,0)$.
   If $b=c+d$, then for $x_0=0$ we have $C=d_0=0$, so the slice is the point $(0,0)$ by (ii).
\end{proof}

When one assumes in the degenerate triangle case $b=c+d$
that $x,y,z$ are small then
\begin{align}
 \det D & = -4cd x^2 + 4d(c+d) y^2 + 4c(c+d) z^2 + \mathcal{O}(\norm{(x,y,z)}^4)
\end{align}
The quadratic part defines the tangent double cone with axis along the $x$-axis at the origin, and it follows that in
the case of 3 solutions, the zero mode surface consists of two topological spheres
which touch each other at the origin, where the surface has a conical singularity.
(For $c=d$ the zero mode surface is rotationally symmetric about the $x$-axis;
for $b=1$, $c=d=\tfrac12$ it is the rotated lemniscate of Bernoulli discussed in \cite{DLS22,CLL26}.)

\medskip
In summary, together with Proposition \ref{prop_slice}, we can formulate the following theorem.

\begin{theorem} \label{theo_m=3}
  Assume $d=0$ and $b\ge c\ge 0$.
  Then the zero mode surface $\det D =0$ consists of the one or two points
  \begin{align}
     \paraa{\pm\sqrt{b^2-c^2}, 0, 0 }
  \end{align}
  Assume $b\ge c\ge d > 0$.
  Then the zero mode surface $\det D =0$ is:
  \begin{enumerate}[label=\textup{(\roman*)}]
    \item two disjoint topological spheres, if $b>c+d$;
    \item two topological spheres touching at the origin in a conical singularity, if $b=c+d$; and
    \item one topological sphere, if $b<c+d$.
  \end{enumerate}
\end{theorem}

\medskip
Heron's formula (\ref{heron}) gives a geometric interpretation of the transition
when $d>0$: a nondegenerate triangle corresponds to one sphere, a degenerate
triangle to two spheres meeting at a point, and failure of the triangle inequality
to two disjoint spheres. When $d=0$, the spectrum instead consists of the one or two
points $(\pm\sqrt{b^2-c^2},0,0)$; they coincide if $b=c$. For $b>c$, their distance
from the origin is the other leg of a right triangle with hypotenuse $b$ and leg $c$.

\medskip
Comparing $R_\pm$ of proposition \ref{prop_slice} for $d>0$ with the
solution (\ref{xpm_0}) for $d=0$, we see that the points
$x_{\pm,d=0}$ are always inside the zero mode surface for $d\neq 0$.
In particular, the points $x_{\pm,d=0} = \paraa{\pm\sqrt{b^2-c^2}, 0, 0 }$
(\ref{xpm_0}) are the zero mode surface $\det D=0$ if $d=0$, and strictly inside it,
i.e.\ in the bounded region where $\det D<0$, if $d>0$.

In the graph construction of \cite{Sykora16}, the normal form with $d=0$
is the basic building block with both nodes at the origin, joined by one edge with string radii $s_x=b$ and $s_y=c$.
The edge is visualized as a closed string that opens into an ellipse with semi-axes $b\ge c$ in the $x$- and $y$-direction.
The points $x_{\pm,d=0}=\paraa{\pm\sqrt{b^2-c^2},0,0}$ are exactly the foci of this ellipse,
and they are the complex eigenvalues of $V=X+iY$.
For $d>0$, the two nodes are separated to $z=\pm d$ and lie strictly inside the zero mode surface.
Thus a graph whose vertices sit at the points (\ref{xpm_0}) is the limit
$d\to0$ of such a building block, in which the surface collapses onto these two points.

\medskip
We now return to an arbitrary number $m$ of matrices $X^a$ .
Together with Proposition \ref{prop_zero_matrix}, Theorem \ref{theo_m=3}
determines the Clifford spectrum of any number of Hermitian $2\times 2$ matrices.

\begin{theorem}\label{theo_all_m}
  Let $X^1,\dots,X^m$ be Hermitian $2\times 2$ matrices and let $b_1\geq b_2\geq b_3\geq 0$ be the
  singular values of the coefficient matrix $M=(x^a_i)$ defined by
  \begin{align}
   X^a = x^a_0 \sigma_0 +x^a_i\sigma_i
\end{align}
 where $b_a=0$ for $a>m$ if $m<3$.
  Up to a rotation and a shift in $\reals^m$, the Clifford spectrum $\Lambda(X^1,\dots,X^m)$ is
  \begin{enumerate}[label=\textup{(\roman*)}]
    \item $\Lambda(b_1\sigma_1,b_2\sigma_2,b_3\sigma_3)\times\{0\}^{m-3}$ for $m\geq 3$, where
      $\Lambda(b_1\sigma_1,b_2\sigma_2,b_3\sigma_3)$ is given by Theorem \ref{theo_m=3};
    \item the one or two points $\paraa{\pm\sqrt{b_1^2-b_2^2},0}$ for $m=2$;
    \item the one or two points $\pm b_1$ for $m=1$.
  \end{enumerate}
\end{theorem}
\begin{proof}
  The first statement follows from the normal form and Proposition \ref{prop_zero_matrix}.
  For $m<3$, Proposition \ref{prop_zero_matrix} implies
  $\Lambda(b_1\sigma_1,\dots,b_m\sigma_m)\times\{0\}^{3-m}=\Lambda(b_1\sigma_1,b_2\sigma_2,0)$ with $b_2=0$
  for $m=1$, which by the theorem consists of the points $\paraa{\pm\sqrt{b_1^2-b_2^2},0,0}$.
\end{proof}

For $m=2$ this is in accordance with the observation of \cite{Sykora16} that the zero mode
surface of two matrices consists of the complex eigenvalues of $V=X^1+iX^2$, which are
$\pm\sqrt{b_1^2-b_2^2}$ for $V=\begin{pmatrix} 0 & b_1+b_2\\ b_1-b_2 & 0\end{pmatrix}$.
For $m=1$ we obtain the eigenvalues $\pm b_1$ of $b_1\sigma_1$.

\subsection{The moduli space}\label{subsec_moduli}

With respect to a scaling $X^a \mapsto \lambda X^a$ for a scaling parameter $\lambda > 0$, 
the determinant $\det D(x)$ scales with a factor $\lambda^{M}$,
when one assumes that also the $x^a$ scale with $\lambda$.
Here, $M$ is the size of the matrix of the Dirac operator, which is
$M = n_m N$, where $n_m$ is the dimension of the $\gamma$-matrices.

This means that when $\det D_{X^a}(x)=0$ at $x^a$, it follows that
$\det D_{\lambda X^a}(\lambda x)=0$. When one scales the $X^a$
by $\lambda$, then also the Clifford spectrum scales by $\lambda$.

With respect to the representational set $b_i \sigma_i$ of Proposition \label{prop_b1b2b3rep},
this means that $\lambda b_i$ results in a joint spectrum scaled by $\lambda$.
We can reduce the moduli space by identifying points $(b_1,b_2,b_3) \sim (\lambda b_1, \lambda b_2, \lambda b_3)$.
This results in a set consisting of the origin and a representing point for every ray
from the origin in the cone (\ref{mod_sp_cone}).
This allows to normalize to $b_1=1$.
\begin{align} 
  1 \geq b_2 \geq b_3 \geq 0, \qquad b_1 = 1
\end{align}
The resulting parameter space of normal-form triples is the triangle
$1\geq b_2\geq b_3\geq0$. (The zero triple $b_1=b_2=b_3=0$ is a separate equivalence class,
whose spectrum is a single point.)

\begin{figure}[ht]
\centering
\begin{tikzpicture}[scale=7]
  \fill[gray!12] (0,0) -- (1,0) -- (0.5,0.5) -- cycle;
  \fill[gray!32] (1,0) -- (1,1) -- (0.5,0.5) -- cycle;
  \draw[->] (0,0) -- (1.18,0) node[below] {$b_2/b_1$};
  \draw[->] (0,0) -- (0,1.1) node[left] {$b_3/b_1$};
  \draw (0,0) -- (1,1) -- (1,0);
  \draw[line width=1.6pt] (0,0) -- (1,0);
  \draw[very thick] (1,0) -- (0.5,0.5);
  \foreach \xx/\lab in {0/0, 0.5/\tfrac12, 1/1} \draw (\xx,-0.012) -- (\xx,0.012) node[below=4pt] {$\lab$};
  \fill (0.5,0.5) circle[radius=0.012] node[above left] {lemniscate};
  \fill (1,1) circle[radius=0.012] node[above right] {round sphere};
  \fill (1,0) circle[radius=0.012];
  \node[right] at (1.01,0.05) {one point};
  \node at (0.45,0.15) {two disjoint spheres};
  \node at (0.79,0.54) {one sphere};
  \node[rotate=-45,font=\small] at (0.685,0.255) {$1=b_2+b_3$};
  \node[font=\small] at (0.5,-0.11) {two points on the bottom edge except at $(1,0)$};
\end{tikzpicture}
\caption{Parameter space of nonzero normal-form triples, further normalized to $b_1=1$.}
\label{fig_moduli}
\end{figure}
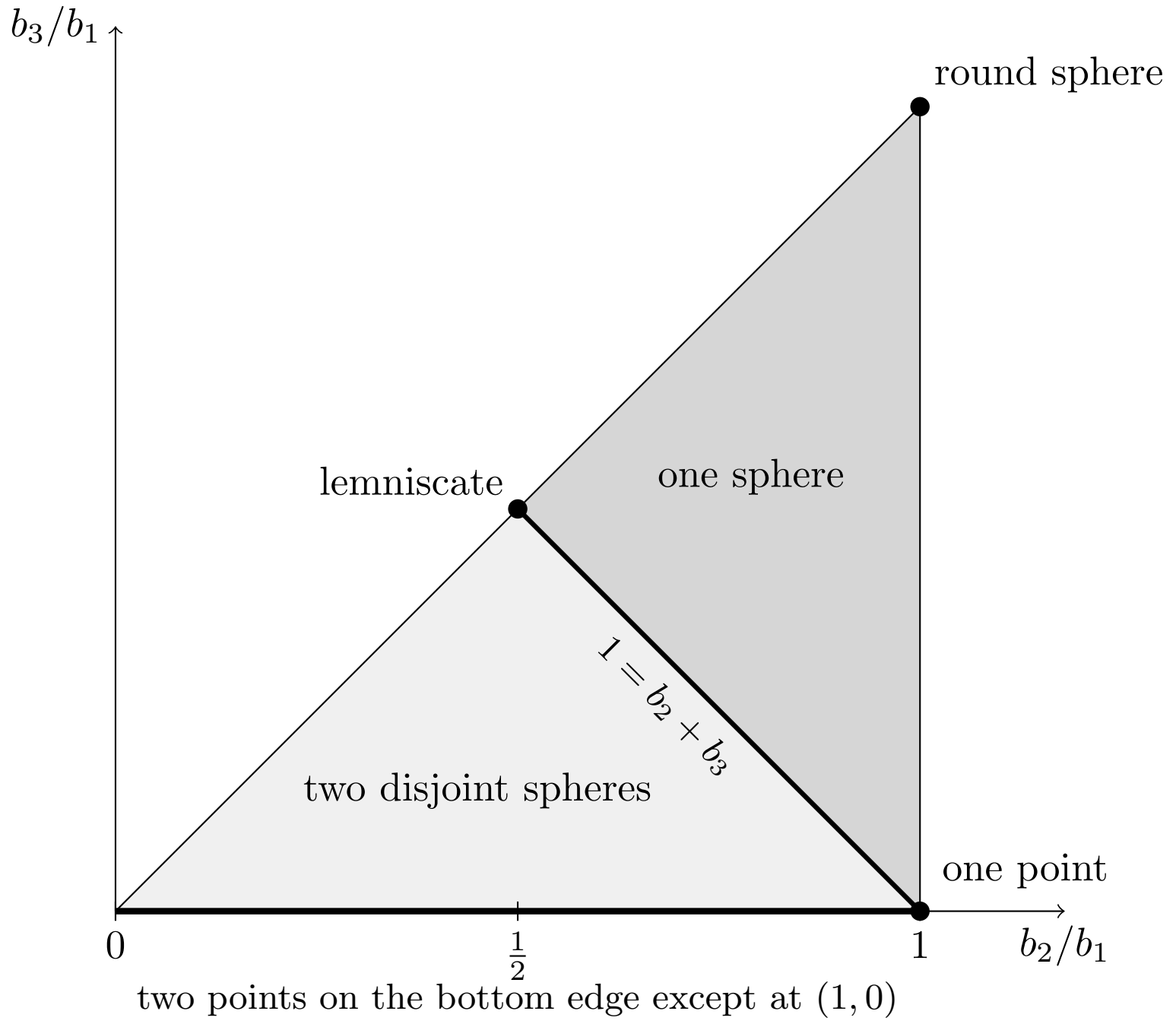

Figure \ref{fig_moduli} indicates the 
topology of the zero mode surface over the normalized triangle. For example, all the
two-point spectra on the bottom edge are equivalent up to translation, rotation
and scaling, although the $X^a$ differ.

The bottom edge $b_3=0$ gives two points except at $(b_2,b_3)=(1,0)$, where they
coincide. The segment $b_2+b_3=1$ with $b_3>0$ gives two touching spheres and
separates the two-sphere and one-sphere regions. Its endpoint $(1,0)$ is excluded
from the touching-spheres classification. 
At $(b_2,b_3)=(1,1)$ the spectrum is a round sphere: restoring the scale gives
$F=(r^2-b_1^2)(r^2+3b_1^2)$, with $r^2=x^2+y^2+z^2$, and hence radius $b_1$.
The point $(\tfrac12,\tfrac12)$ represents the rotated lemniscate of
\cite{DLS22,CLL26}.

\section{Conclusions and outlook}\label{sec_concl}

The Clifford spectrum of a set of $m$ Hermitian $2\times2$ matrices can be
classified in terms of the singular values of
its traceless coefficient matrix $x_i^a$ (\ref{pauli_decomp}).
This gives an invariant parametrization and
a direct topological proof of the configurations
described geometrically in \cite{CLL26}. For three positive singular values, the
triangle-inequality criterion separates one sphere, a conical pinch joining two
spheres, and two disjoint spheres. Rank-deficient cases give one or two points.

Since every triple of Hermitian $2\times 2$ matrices can be transformed into the normal
form, the classification also covers all deformations of the smallest fuzzy sphere
$X^a=\tfrac12\sigma_a$, for example the gauge deformations $X^a\mapsto X^a+A^a$ studied
numerically in \cite{SS24}. In terms of the graphs of \cite{Sykora16}, the result
describes the zero mode surface of the basic building block, i.e.\ of a single edge with
arbitrary labels, and explains in which way such a building block is either connected or
splits into two components.

Several questions remain open. For $N>2$ there is no analogous normal form,
since the adjoint action of $U(N)$ on the $(N^2-1)$-dimensional space of traceless
Hermitian matrices is no longer the full rotation group $SO(N^2-1)$ for $N>2$, and the
moduli space becomes much larger. Nevertheless,
the graph construction of \cite{Sykora16,SS24} suggests that the zero mode surfaces of
larger matrices can be understood locally in terms of $2\times 2$ building blocks, for
example when two zero mode surfaces merge or split \cite{SS24}. It would be interesting
to make this precise and to use it for an approach to the conjecture that the Clifford
spectrum is never empty \cite{CLL26}. Finally, in the context of the spectral localizer
\cite{Loring15,CL22}, the transitions found here are points where the topology of the
Clifford spectrum changes, and it would be interesting to relate them to changes of local
topological invariants.

\section*{Acknowledgements}
The author thanks Paul Schreivogl and Richard Schweiger for useful discussions.

\end{document}